\documentclass[doublecol]{epl2}

\usepackage{amssymb,amsmath}

\usepackage{epsfig}

\usepackage{graphics,graphicx}% Include figure files
\usepackage{color}

\newcommand{\vicente}[1]{{ #1}}

\begin{document}
\title{Non-equilibrium phase transition in a sheared granular mixture}
\author{Vicente Garz\'{o}\inst{1} and Emmanuel Trizac\inst{2}}
\shortauthor{Garz\'{o} and Trizac}

\institute{
  \inst{1}Departamento de F\'{\i}sica, Universidad de Extremadura,
E-06071 Badajoz, Spain\\
  \inst{2}Universit\'e Paris-Sud, Laboratoire de Physique
Th\'eorique et Mod\`eles Statistiques (CNRS UMR 8626), B$\hat{a}$timent 100,
91405 Orsay cedex, France}

\abstract{
The dynamics of an impurity (or tracer particle) immersed in a dilute granular gas under uniform
shear flow is investigated. A non-equilibrium phase transition is identified from an exact solution of
the inelastic Boltzmann equation for a granular binary mixture in the tracer limit, where
the impurity carries either a vanishing (disordered phase)
or a finite (ordered phase) fraction of the total kinetic energy of the
system. In the disordered phase, the granular temperature
ratio (impurity ``temperature'' over
that of the host fluid) is finite, while it
diverges in the ordered phase. To correctly capture
this extreme violation of energy equipartition, we show that the picture
of an impurity enslaved to the host fluid is insufficient.}

%\pacs{05.20.Dd, 45.70.Mg, 51.10.+y}
%\date{\today}

%\begin{document}
\maketitle

Under external driving, an initially mixed macroscopic
granular system may segregate. This ubiquitous
effect is not always desirable in applications, and among the
numerous counter-intuitive properties of granular materials, it has
arguably been among the most studied in the last twenty years
\cite{Seg,Aranson}.
Yet, it is not well understood. The reason is twofold: there is a
large number of {\it a priori} relevant parameters involved in the
description of the granular mixture; additionally, analytical results
are scarce and difficult to obtain. Both shortcomings call
for an improved fundamental understanding of physical situations
where the different physical effects at work (size asymmetry, forcing,
collisional dissipation, etc.) can be deciphered.
As a consequence, seeking
for theoretical progress, we simplify the problem in two respects.
First, we consider the impurity limit where a tracer particle is immersed
in a driven {\em dilute} granular gas described by the inelastic Boltzmann
equation. Second, we focus on an inelastic version of
the so-called Maxwell molecules \cite{E81},
that can be seen as defining the kinetic theorist's Ising model.
Indeed, the Boltzmann equation for elastic bodies is already notoriously
difficult to deal with, and inclusion of collisional dissipation --an essential
aspect of inter-grains collisions-- leads to a far more complex description.
The corresponding inelastic Maxwell model, that has witnessed an
upsurge of interest in recent years
\cite{B02,E02,E06,SG07,A10,IMM}, is inspired by Maxwell's original insight \cite{Maxwell}
that scattering processes
involving collision rates independent of the relative velocity of impacting
particles, define mathematically tractable approaches \cite{rque15}.

A valid question to raise from the outset
is then that of the relevance of such a simplification. This
issue should
be addressed by comparison with more refined approaches. Among the
few well accepted models accounting for the specifics of
dissipative inter-grains collisions, the simplest is
the inelastic hard sphere model \cite{Aranson,BP}, where results
pertaining to the inelastic Boltzmann equation can
be compared to their Maxwell model counterpart. While the agreement
concerning the Navier-Stokes transport coefficients
is only qualitative \cite{S03}, the situation improves significantly for sheared
mixtures where the rheological properties (shear and normal
stresses) agree very well with (approximate) analytical results and computer simulations
for inelastic hard spheres \cite{G03}.
This agreement, in a parameter
space of large dimensionality, gauges and establishes
the reliability of the Maxwell
model to capture important effects in \emph{sheared}
granular mixtures. We come back to the relevance
of our approach
before our concluding section, see below,
%for sheared granular mixture, the situation that will be
%addressed in the subsequent analysis.
and we further note that experiments with magnetic grains also point to
the usefulness of Maxwell models \cite{K05}.

Our objective is to investigate the dynamics of an impurity or tracer particle (hereafter referred to
as species 1) in an inelastic dilute gas (species 2),
maintained in a non-equilibrium steady state by a uniform shear flow.
This particular state is amenable to analytical treatment
from the Boltzmann equation since it is characterized by constant and uniform partial
densities ($n_i$ for $i=1,2$) as well as a uniform
``granular temperature'' $T(t)$ \cite{rque16}, while the only
gradient affects the velocity profile, that is furthermore
common to both species: $u_x = a y$ in a Cartesian frame
[$(x,y,z)$ in three dimensions].
This linear velocity field defines
the constant shear rate $a$ that quantifies viscous heating (shearing work).
This energy injection mechanism is balanced by dissipation in inelastic
collisions, the latter being parameterized by the constant coefficients
of normal restitution $\alpha_{ij}$ for collisions between species $i$
with $j$ \cite{BP}.
Although the total temperature $T=(n_1 T_1+n_2T_2)/(n_1+n_2)$
($T_i$ is the partial temperature of species $i$)
changes with time, the temperature ratio $T_1/T_2$ is time independent in the hydrodynamic
regime (\vicente{namely, for times much larger than the mean free time)}. Note that equipartition, of course, has no reason
to hold in such non equilibrium conditions: $T_1/T_2\neq 1$ \cite{wp02}.
A plausible expectation is that upon taking the tracer
limit $c_1\to 0$ (where $c_i=n_i/(n_1+n_2)$ is the mole fraction of species $i$),
\vicente{the properties of the medium (or granular gas) are not affected by the presence of the tracer particles (or impurity). Let us consider the relative contribution of the tracer particles to the total energy of the system, $E_1/E=c_1T_1/T$. Out of equilibrium, one expects that} $E_1/E$ actually
scales like $c_1$, and should
therefore vanish \vicente{when $c_1\to 0$}. In other words, the impurity should be enslaved to
the host medium, and should not \vicente{change} the macroscopic properties of the
system. \vicente{In this letter we present an example of a violation of the above expectation:} A non-equilibrium
phase transition takes place, that allows to discriminate
a \emph{disordered} phase, where $T_1/T$ in the tracer limit is finite
--and hence $E_1/E=0$--, from an \emph{ordered} phase where $E_1/E$ is finite
--and therefore $T_1/T$ diverges. The corresponding regions of phase
space will be worked out explicitly, which will reveal that both
phases may exhibit unexpected reentrant features.
The apparition of an
ordered phase corresponds to an extreme violation of equipartition,
and has already been reported, but only in the case of an unforced system ($a=0$)
\cite{NK02}. However, this work was performed at the ``enslaved impurity''
level, neglecting the retroaction of impurity onto the host fluid.
This precludes the derivation of the order parameter.
Compared to the findings of Ref.\ \cite{NK02}, we obtain explicitly
the value of the transition order parameter $E_1/E$ in the ordered phase,
and show that the scenario is more complex than reported so far,
with the emergence of overlooked phases.
Furthermore, we investigate the fate of the previous phases, including
the new ones obtained, at finite shear ($a\neq 0$).
These new results are the most significant contribution of the present
work.

\vicente{We consider a binary mixture of grains at low density,
driven by a uniform shear flow. In the Lagrangian frame moving at the linear flow velocity ${\bf u}$, the velocity distribution functions $f_i ({\bf r}, {\bf v},t)$ of each species become uniform, i.e., $f_i({\bf r}, {\bf v},t)\equiv f_i({\bf V},t)$ where ${\bf V}={\bf v}-{\bf u}$. In this frame, the set of coupled Boltzmann equations for the velocity distributions $f_i({\bf V},t)$ read \cite{GS03}
\begin{equation}
\label{n1} \frac{\partial}{\partial t}f_i-aV_y\frac{\partial}{\partial
V_x}f_i=\sum_j\,J_{ij}[{\bf v}|f_i,f_j],
\end{equation}
where the Boltzmann collision operator $J_{ij}[f_i,f_j]$ is
\begin{eqnarray}
&& J_{ij}\left[{\bf V}_{1}|f_{i},f_{j}\right] =\frac{\omega_{ij}}{n_j\Omega_d}
\int d{\bf V}_{2}\int d\widehat{\boldsymbol {\sigma }}\left[ \alpha_{ij}^{-1}f_{i}({\bf V}_{1}')f_{j}(
{\bf V}_{2}')\right.\nonumber\\
& & \left.-f_{i}({\bf V}_{1})f_{j}({\bf V}_{2})\right]
\;.
\label{n2}
\end{eqnarray}
Here, $\Omega_d=2\pi^{d/2}/\Gamma(d/2)$ is the total solid angle in $d$ dimensions, and $\alpha_{ij}\leq 1$ denotes the
(constant) coefficient of restitution  for collisions between particles of species $i$
with $j$. Moreover, ${\bf V}_{1}'={\bf V}_{1}-\mu_{ji}\left( 1+\alpha_{ij}
^{-1}\right)(\widehat{\boldsymbol {\sigma}}\cdot {\bf g}_{12})\widehat{\boldsymbol
{\sigma}}$, ${\bf V}_{2}'={\bf V}_{2}+\mu_{ij}\left(
1+\alpha_{ij}^{-1}\right) (\widehat{\boldsymbol {\sigma}}\cdot {\bf
g}_{12})\widehat{\boldsymbol{\sigma}}$,
where ${\bf g}_{12}={\bf V}_1-{\bf V}_2$, $\widehat{\boldsymbol {\sigma}}$ is a unit vector directed along the centers of the two colliding spheres, and $\mu_{ij}=m_i/(m_i+m_j)$. The effective collision frequencies $\omega_{ij}$ for collisions
$i$-$j$ are independent of the relative velocities of the colliding particles but can depend on space and time through its dependence on densities $n_i$ and temperature $T$. They can be also seen as free parameters of the model. Here, since our problem involves a delicate tracer limit,}
we aim at the simplest possible approach: In the ``plain vanilla''
Maxwell model worked out here, the collision frequencies $\omega_{ij}$  are of the form
$\omega_{ij} = \nu c_j$, where $\nu^{-1}$ is an effective (constant) mean free time.
\vicente{The form of $\omega_{ij}$ is closer to the original model of Maxwell molecules for ordinary gas mixtures \cite{Maxwell}. The plain vanilla Maxwell model has been previously considered by several authors \cite{NK02,vanilla} in several problems pertaining to granular mixtures.}  Other choices of $\omega_{ij}$ are possible,
e.g. if the goal lies in mimicking as accurately as possible inelastic
hard spheres \cite{G03}, but at the price of a substantial increase in the
complexity of the model that prevents to get exact results.

\vicente{In the uniform shear flow problem,} the
reduced shear rate $a^* = a/\nu$ is the relevant nonequilibrium parameter since it measures the distance
from the unforced case, the
much studied so-called homogeneous cooling state (HCS) \cite{BP}.
Within our framework, $a^*$ does not depend on time, so
that in general $a^*$ and $\alpha_{ij}$ are independent parameters. This
decoupling
allows us to analyze the combined effect of both control parameters on the
dynamic properties of the granular mixture. 
\vicente{This is one of the main advantages of the vanilla model used here in contrast to previous works \cite{G03}.}
More generally, a key feature of the Boltzmann equation for Maxwell models is
that the pressure tensor ${\sf P}$  can be exactly determined in terms of the
shear rate and the parameters of the mixture (the mass ratio $\mu\equiv m_1/m_2$, the mole
fraction $c_1=n_1/(n_1+n_2)$ and the coefficients of restitution $\alpha_{ij}$)
\cite{G03,GT10}. \vicente{The knowledge of the pressure tensor allows one to determine the non-Newtonian transport properties of the mixture. The energy ratio $E_1/E$ is also an important property that can be obtained from the pressure tensor.}

\vicente{We now address the tracer limit ($c_1\to 0$) for $E_1/E$.}
After some tedious but straightforward algebra \cite{GT11}, an interesting
phenomenon is uncovered in this limit: while $E_1/E$ vanishes
for some parameter range --this is the naive expectation--, the possibility
exists that $E_1/E \neq 0$. Two relaxation rates, $\lambda_1^{(0)}$ and
$\lambda_2^{(0)}$, play a key role in delimitating the two behaviours
\cite{GT11}:
whenever $\lambda_2^{(0)}>\lambda_1^{(0)}$, one has $E_1/E=0$ while
on the other hand, $E_1/E \neq 0$ for $\lambda_2^{(0)}<\lambda_1^{(0)}$.
The corresponding expressions for $\lambda_2^{(0)}$ and $\lambda_1^{(0)}$ are \cite{GT11}
\begin{equation}
\label{2}
\lambda_2^{(0)}=\frac{(1+\alpha_{22})^2}{d+2}\varphi(\widetilde{a})-\frac{1-\alpha_{22}^2}{2d},
\end{equation}
\begin{eqnarray}
\label{2.1}
\lambda_1^{(0)} &=&\frac{2\mu_{21}^2}{d+2}(1+\alpha_{12})^2\varphi
\left(\frac{\widetilde{a}}{2\mu_{21}^2}
\left(\frac{1+\alpha_{22}}{1+\alpha_{12}}\right)^2\right) \nonumber\\
&&-\frac{2}{d}\mu_{21}(1+\alpha_{12})\left[1-\frac{\mu_{21}}{2}(1+\alpha_{12})
\right],
\end{eqnarray}
where $\widetilde{a}=2(d+2)a^*/(1+\alpha_{22})^2$,
and $\varphi(x)\equiv \frac{2}{3}\sinh^2[\frac{1}{6}\cosh^{-1}(1+\frac{27}{d}x^2)]$.
In cases where $\lambda_1^{(0)}>\lambda_2^{(0)}$, \vicente{the expression of $E_1/E$ can be written as
\begin{equation}
\label{n3}
\frac{E_1}{E}=\frac{D(\lambda_1^{(0)})}{\Delta_{01}(\lambda_1^{(0)})\lambda_1^{(1)}+
\Delta_1(\lambda_1^{(0)})},
\end{equation}
where the dependence of $m_1/m_2$, $\alpha_{ij}$, and $a^*$ is implicitly assumed on the right-hand side. In addition, $\lambda_1^{(1)}$ is defined by the expansion
$\lambda_1(a^*,c_1)\approx \lambda_1^{(0)}(a^*)+\lambda_1^{(1)}(a^*) c_1+{\cal
O}(c_1^2)$ where $\lambda_1$ is the largest real root of a sixth-degree polynomial equation \cite{GT11}. The general expressions of $D$, $\Delta_{01}$ and $\Delta_1$ are too lengthy to be written down here. For the sake of illustration, we give their expressions for the case $\frac{m_1}{m_2}=\frac{1}{4}$ and
$\alpha_{11}=\alpha_{22}=\alpha_{12}=\frac{1}{2}$:
%\begin{widetext}
\begin{eqnarray}
\label{n4}
& & D=-\frac{9}{25}\left[\lambda_1^{(0)}+\frac{7}{20}\right]^2\left[\lambda_1^{(0)}+\frac{76}{125}\right]^2
\nonumber\\
& &
-\frac{18a^{*2}}{125}\left[\lambda_1^{(0)}(1+\lambda_1^{(0)})+\frac{1331}{5000}\right],
\end{eqnarray}
\begin{eqnarray}
\label{n5}
&&\Delta_{01}
=\frac{144}{125}
\left(\lambda_1^{(0)}+\frac{7}{20}\right)\left\{
a^{*2}-\frac{625}{48}\left(\frac{76}{125}+\lambda_1^{(0)}\right)
\right.\nonumber\\
& & \times\left.
\left[\frac{2581}{15625}+\lambda_1^{(0)}\left(\frac{2077}{2500}+\lambda_1^{(0)}\right)\right]
\right\},
\end{eqnarray}
\begin{eqnarray}
\label{n6}
& & \Delta_1
=
\frac{153a^{*2}}{500}\left(\lambda_1^{(0)2}+\frac{223}{170}\lambda_1^{(0)}
+\frac{55857}{170000}\right)
\nonumber\\
& & -\frac{75831}{125000}\left(\lambda_1^{(0)}+\frac{7}{20}\right)\left(\lambda_1^{(0)}+\frac{76}{125}\right)
\nonumber\\
& & \times
\left[1+5\lambda_1^{(0)}\frac{10403+132100 \lambda_1^{(0)}}{101108}\right].
\end{eqnarray}
It must be noted that in general the expression of $E_1/E$ does not explicitly depend on the coefficient of restitution $\alpha_{11}$ 
which is equivalent to neglecting the collisions among tracer particles themselves in the kinetic equation of $f_1$.}

The change of behaviour (energy ratio that vanishes or not)
is akin to an ordering process where the tracer is either enslaved to the
host fluid ($E_1/E = 0$), or carries a finite fraction of the total
kinetic energy of the system ($E_1/E \neq 0$). The latter situation,
where the temperature ratio is divergent, will be subsequently referred
to as the ``ordered phase'', characterized by an extreme breakdown
of energy equipartition. Loosely speaking, the system is invariant
under the transformation $c_1\to \delta c_1$ in the disordered phase, where
$T_1/T$ reaches a finite value in the limit $c_1\to 0$. This invariance
is broken in the ordered phase, since then $T_1/T \propto c_1^{-1}$.

In light of the previous discussion, it is instructive to analyze some
special cases. First,
for sheared elastic gases ($a^*\neq 0$ and $\alpha_{12}=\alpha_{22}=1$), it
is easy to see that $\lambda_2^{(0)}>\lambda_1^{(0)}$ (``disordered'' phase)
if the mass ratio $\mu>\sqrt{2}-1\simeq 0.414$ for any value of $a^*$. However, if $\mu<\sqrt 2-1$,
$\lambda_1^{(0)}>\lambda_2^{(0)}$ (``ordered'' phase) for $a^*$ larger than a critical value $a_c^*(\mu)$.
We recover here the transition already found \cite{MSG96}
for elastic Maxwell mixtures under uniform shear flow.
A second interesting situation is that of the HCS ($a^*=0$ but
$\alpha_{ij}\neq 1$). In this case, there are
also two different phases depending on the relative positions of
$\lambda_1^{(0)}$ and $\lambda_2^{(0)}$. \vicente{In particular,} the transition point
($\lambda_1^{(0)}=\lambda_2^{(0)}$) --where it can be shown that the
system relaxation time diverges-- leads
to the critical mass ratios $\mu_{\text{HCS}}^{(\pm)}=(\alpha_{12}\pm
\sqrt{(1+\alpha_{22}^2)/2})/(1\mp\sqrt{(1+\alpha_{22}^2)/2}$ with $\mu_{\text{HCS}}^{(-)}<\mu_{\text{HCS}}^{(+)}$.
The ordered phase ($\lambda_1^{(0)}>\lambda_2^{(0)}$)
exists for mass ratios smaller
than $\mu_{\text{HCS}}^{(-)}$ or larger than $\mu_{\text{HCS}}^{(+)}$.
Moreover, the first ordered phase ($\mu<\mu_{\text{HCS}}^{(-)}$
corresponding to light impurities)
is present only when $\alpha_{12}>\sqrt{(1+\alpha_{22}^2)/2}$, and so is absent when
$\alpha_{12}=\alpha_{22}$ or when $\alpha_{12}<1/\sqrt{2}$.
It therefore appears that the occurrence of this ordered phase
not only requires collisional dissipation but more precisely, asymmetric dissipation.
\vicente{In the absence of shear ($a^*=0$), the order parameter $E_1/E$ can be cast in a simple form as}
\begin{equation}
\label{10}
\frac{E_1}{E} =
\frac{\alpha_{22}^2-1+4\mu_{21}(1+\alpha_{12})\left[1-\frac{\mu_{21}}{2}(1+\alpha_{12})\right]}
{\alpha_{22}^2-1+2\mu_{21}(1-\alpha_{12}^2)}.
\end{equation}
\begin{figure}[htb]
\begin{center}
\includegraphics[width=0.45\textwidth]{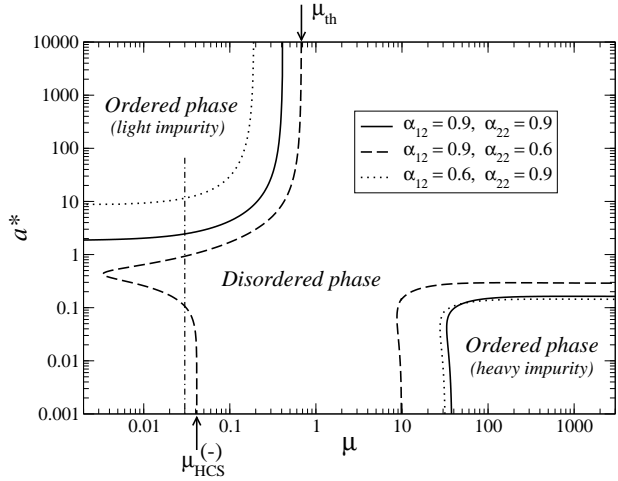}
\caption{Phase diagram in the reduced shear rate vs mass ratio plane, for different
coefficients of restitution. The lines display the boundaries between
ordered and disordered phases;
the light and heavy impurity ordered phases
are indicated.
The vertical dot-dashed line corresponds to a
cut leading to the inset of Fig. \ref{fig:2}. The arrows shows
$\mu_{\text{HCS}}^{(-)}$ and $\mu_{\text{th}}$
for $\alpha_{12}=0.9$ and $\alpha_{22}=0.6$.
For the two other parameters sets,
\{$\alpha_{12},\alpha_{22}\}=\{0.9,0.9\}$ and $\{0.6,0.9\}$,
the thresholds $\mu_{\text{th}}$ and $\mu_{\text{HCS}}^{(+)}$
exist (not shown, for clarity) while $\mu_{\text{HCS}}^{(-)}$
is not defined. Here, $d=2$, but very similar results are
obtained in three dimensions.
\label{fig:1}}
\end{center}
\end{figure}
\begin{figure}[tb]
\begin{center}
\includegraphics[width=0.45\textwidth]{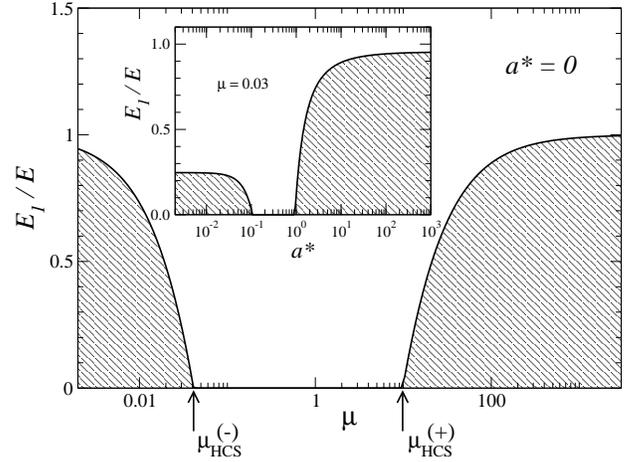}
\caption{Order parameter of the non equilibrium transition,
as a function of mass ratio $\mu = m_1/m_2$ in the unforced case
for a two-dimensional system.
Here, $\alpha_{12}=0.9$ and $\alpha_{22}=0.6$
The inset shows the same quantity as a function of reduced shear rate,
for $\mu=0.03<\mu_{\text{HCS}}^{(-)}$.
The hatched regions indicate the ordered phases
(reentrant light impurity phase in the inset).
The two thresholds in the unforced case are indicated by the vertical arrows
($\mu_{\text{HCS}}^{(-)}\simeq 0.041$ and $\mu_{\text{HCS}}^{(+)}\simeq 9.83$).
 \label{fig:2}}
\end{center}
\end{figure}
%and $\mu_{\text{HCS}}^{(+)}=(\alpha_{12}+\sqrt{(1+\alpha_{22}^2)/2})/(1-\sqrt{(1+\alpha_{22}^2)/2}$,
 %\begin{equation}
%\label{3}
%\mu_{\text{HCS}}^{(-)}=\frac{\alpha_{12}-\sqrt{\frac{1+\alpha_{22}^2}{2}}}
%{1+\sqrt{\frac{1+\alpha_{22}^2}{2}}}, \quad
%\mu_{\text{HCS}}^{(+)}=\frac{\alpha_{12}+\sqrt{\frac{1+\alpha_{22}^2}{2}}}
%{1-\sqrt{\frac{1+\alpha_{22}^2}{2}}},
%\end{equation}
The existence of the second ordered phase ($\mu>\mu_{\text{HCS}}^{(+)}$,
heavy impurities)
was already found by Ben-Naim and Krapivsky \cite{NK02} in their analysis
on the velocity statistics of an impurity immersed in a uniform granular fluid.
It must be remarked that a similar
extreme breakdown of the energy equipartition has also been
reported \vicente{in the HCS} for inelastic hard spheres \cite{SD01} since, in the ordered phase,
the ratio of the mean square velocities for the impurity and fluid particles
$T_1m_2/T_2m_1$ is finite for
extremely large mass ratios ($m_1/m_2\to \infty$).

We now turn to the general case where \vicente{viscous heating} and collisional dissipation are both
at work. Unlike with the HCS where
the light impurity ordering requires asymmetric dissipation,
we find here that adding the ``shear dimension'' to the phase
diagram ensures the systematic existence of the light impurity
ordered phase. This is illustrated in Fig. \ref{fig:1}, see the left
hand side. On the other hand, the heavy impurity order sets in
at large enough mass ratio and small enough shear (lower right corner
of Fig. \ref{fig:1}).
As anticipated from the study of the unforced system, the behaviour
differs depending if the inequality $\alpha_{12}>\sqrt{(1+\alpha_{22}^2)/2}$
holds or not. In Fig. \ref{fig:1}, the parameter set fulfilling the above constraint is
$\alpha_{12}=0.9$ and $\alpha_{22}=0.6$. In the low shear rate
limit, the figure indeed exhibits two distinct ordered phases, one at low
mass ratio $\mu$, and one in the opposite heavy intruder limit. This
behaviour is fully consistent with the previously analyzed HCS scenario. Conversely, when
$\alpha_{12}<\sqrt{(1+\alpha_{22}^2)/2}$, the light impurity ordered
pocket is more restricted, and confined in a portion of high shear,
$a^*>a_c^*$, with quite a complex dependence on the coefficients of restitution $\alpha_{22}$ and $\alpha_{12}$ (see fig. \ref{fig:1}).
In all cases, the light impurity phase can only exist provided
$\mu$ does not exceed some dissipation dependent threshold,
$\mu< \mu_{\text{th}}$, with $\mu_{\text{th}}=\sqrt{2}(1+\alpha_{12})/(1+\alpha_{22})-1$.
%\begin{equation}
%\mu_{\text{th}}=\sqrt{2} \, \frac{1+\alpha_{12}}{1+\alpha_{22}}-1.
%\end{equation}
As can be seen in Fig. \ref{fig:1}, the order/disorder threshold
shear rate diverges as $\mu \to \mu_{\text{th}}$.
Of course, $\mu_{\text{th}}=\sqrt{2}-1$ for elastic gases,
as discussed above.
The phase
diagram shown in Fig. \ref{fig:1} also indicates that the ordered phase
exhibits reentrant features. Upon increasing mass ratio at fixed
reduced shear rate, or conversely increasing shear rate at fixed mass ratio,
the following sequence may be observed for some portion of phase space:
a first transition from ordered to disordered states, followed by
a reverse disorder $\to$ order transition. This is illustrated in
Fig. \ref{fig:2}, where the fraction of the total energy that is transported
by the impurity is plotted against either mass ratio, or shear rate. \vicente{It is apparent that,
for asymptotically large shear rates, the tracer contribution to the total energy can be even larger than that of the excess component.}
\begin{figure}[tb]
\begin{center}
\includegraphics[width=0.45\textwidth]{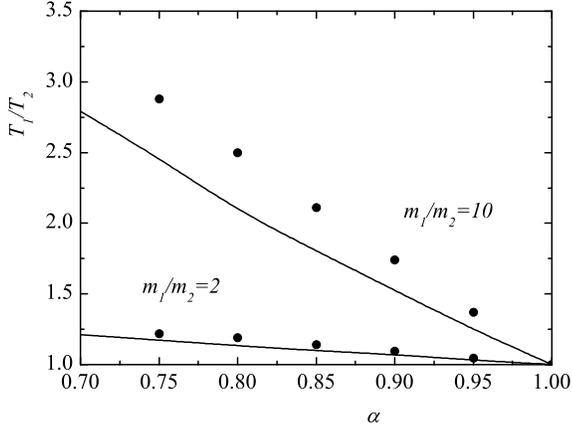}
\caption{Temperature ratio $T_1/T_2$ versus the (common) coefficient of restitution
($\alpha=\alpha_{11}=\alpha_{12}=\alpha_{22}$) in the \emph{steady} uniform shear flow state.
The predictions of the inelastic Maxwell model
(continuous curve,
present work)
are tested against Monte Carlo data (symbols,
from \cite{MG02}).
The parameters are $c_1=0.5$ (equimolar mixture) for a three dimensional system
($d=3$). Two mass
ratios are considered, corresponding to
moderately heavy intruders.
 \label{fig:comp}}
\end{center}
\end{figure}

Before concluding, we come back to the reliability of our ``plain vanilla''
Maxwell approach. Although approximate, it
turns out to provide a relevant framework for
granular binary mixtures under uniform shear flow. This is
illustrated in fig.\ \ref{fig:comp} \vicente{for an equimolar mixture in the steady uniform shear flow state, namely, when viscous heating and collisional cooling cancel each other and consequently, the (reduced) shear rate $a^*$ depends on the coefficients of restitution $\alpha_{ij}$.} Figure \ref{fig:comp} compares
the predictions of the plain inelastic Maxwell model
to the  Monte Carlo results obtained within the more realistic
inelastic hard sphere model.
It can be seen that the plain Maxwell model prediction captures
the important trend observed within a more refined framework (\vicente{the discrepancies between theory and simulation are less than 15\%, even for strong values of dissipation).}
We therefore expect that the transition found in this paper
is not artefactual but is a robust feature.

\vicente{In conclusion, we have analysed the tracer limit of an \emph{exact} solution \cite{G03,GT10} of the Boltzmann equation for a binary granular mixture of inelastic Maxwell gases. This solution applies to \emph{arbitrary} values of the shear rate $a$ and the parameters of the mixture, namely, the mole fraction $c_1$, the mass ratio $\mu$, and the coefficients of restitution $\alpha_{ij}$. We have argued
that when the granular system is driven by an externally imposed (uniform) shear,
the inelastic Maxwell model correctly encodes the physics
of more complex models, as evidenced by the dependence of the non-Newtonian transport
properties on the numerous parameters of the mixture \cite{G03}.

Within the framework of the mean-field Maxwell model, we have shown
a quite \emph{unexpected} result: the relative contribution of the tracer species (or impurity) to the total properties of the mixture does not necessarily tend to zero as $c_1\to 0$. Consequently,
the seemingly natural ``enslaved impurity'' picture \cite{rque20} breaks down. This surprising result extends to \emph{granular} gases some related results derived time ago for ordinary sheared mixtures \cite{MSG96}. The phenomenon discovered here, arising from an exact solution of the
set of coupled Boltzmann equations, has been illustrated with the energy
ratio $E_1/E$ as a probe, but identical conclusions can be drawn
for other properties of the system such as the (intrinsic) shear viscosity and the viscometric functions \cite{Y71}.

The corresponding extreme kinetic energy equipartition violation can be seen as an ordering
transition, governed by the competition between two characteristic
relaxation frequencies, $\lambda_1^{(0)}$ and $\lambda_2^{(0)}$. We
have found two different classes of ordered phases (see the shear rate versus mass ratio diagram
in fig. \ref{fig:1}): a \emph{light} impurity phase that exists when $\mu<\mu_{\text{HCS}}^{(-)}$ for shear rates larger than a certain critical value and a \emph{heavy} impurity phase that requires
$\mu>\mu_{\text{HCS}}^{(+)}$ and shear rates smaller than a certain threshold value. As fig.\ \ref{fig:2} clearly shows, both ordered phases exhibit reentrant features.

Usually, in the tracer limit, one assumes that the state of the excess component $2$ (or
granular gas) is not disturbed by collisions with tracer particles $1$ (or impurity) and that
the self-collisions among tracer particles can be also neglected \cite{RL77}. The results
derived in this letter show that while the second hypothesis is correct, the first expectation
fails in the ordered phase. In this sense, the tracer limit must be taken with care since it
presents, in the ordered phase, a similar complexity as the general problem with finite concentration. In addition, although 
our present description has focused on the energy ratio for a binary system, we expect 
that the results reported here should have interesting consequences for polydisperse mixtures, or in terms of the velocity
statistics, not addressed here. Finally, an important simplification allowed
by our modelisation lies in the decoupling between shear rate and dissipation,
while granular gases exhibit --in the steady state--
inherent coupling between inelasticity and
spatial gradients. This means that a fingerprint
of our scenario should be sought for experimentally in the hydrodynamic transient
regime, before the steady state where
collisional dissipation and viscous heating balance each other. 
We hope that this letter stimulates the performance of such experiments and/or computer
simulations to detect the transition phenomenon reported here.}

\acknowledgments
The research of V.G. has been supported by
the Ministerio de Ciencia e Innovaci\'on  (Spain) through grant No. FIS2010-16587,
partially financed by FEDER funds and by the Junta de Extremadura (Spain) through Grant No. GRU10158.


\begin{thebibliography} {99}


\bibitem{Seg}
See e.g. KUDROLLI A., \emph{Rep. Prog. Phys.}, {\bf 67} (2004) 209
and references therein. For more recent contributions, see for instance, CIAMARRA M. P., CONIGLIO A. and NICODEMI M., \emph{Phys. Rev. Lett.}, {\bf 94} (2005) 188001; SCHNAUTZ T., BRITO R., KRUELLE C. A. and REHBERG I., \emph{Phys. Rev. Lett.}, {95} (2005) 028001; BREY J. J., RUIZ-MONTERO M. J. and
MORENO F., \emph{Phys. Rev. Lett.}, {\bf 95} (2005) 098001 (2005); S\'ANCHEZ I., GUTI\'ERREZ G., ZURIGUEL I. and MAZA D., \emph{Phys. Rev. E}, {\bf 81} (2010) 062301; CLEMENT C.P., PACHECO-MARTINEZ H.A., SWIFT M.R. and KING P. J., \emph{Europhys. Lett.}, {\bf 91} (2010) 54001.

\bibitem{Aranson}
ARANSON I.S. and TSIMRING L.S., \emph{Rev. Mod. Phys.}, {\bf 78} (2006) 641.

\bibitem{E81}
ERNST M. H., \emph{Phys. Rep.}, {\bf 78} (1981) 1.

\bibitem{B02}
BALDASARRI A., MARINI BETTOLO MARCONI U. and PUGLISI A.,
\emph{Europhys. Lett.}, {\bf 58} (2002) 14.

\bibitem{E02}
ERNST M.H. and BRITO R.,
\emph{Europhys. Lett.}, {\bf 58} (2002) 182.


\bibitem{E06}
ERNST M.H., TRIZAC E. and BARRAT A.,
\emph{Europhys. Lett.}, {\bf 76} (2006) 56.

\bibitem{SG07}SANTOS A. and GARZ\'O V., \emph{J. Stat. Mech.}, (2007) P08021.

\bibitem{A10}
ALASTUEY A. and PIASECKI J.,
\emph{J. Stat. Phys.}, {\bf 139} (2010) 991.


\bibitem{IMM}
For some recent reviews see e.g.  BEN-NAIM E. and KRAPIVSKY P. L.,
  \Book{Granular Gas Dynamics}
  \Editor{T. P\"oschel and N. Brilliantov}
  \Vol{624}
  \Publ{Lectures Notes in Physics, Springer, Berlin}
  \Year{2003}
  \Page{65}; VILLANI C.,\emph{J. Stat. Phys.}, {\bf 124} (2006) 781 (2006).



\bibitem{Maxwell}
MAXWELL J.C.,
\emph{Phil. Trans. Roy. Soc.}, {\bf 157} (1867) 49.

\bibitem{rque15}
The kinetic theory for inelastic Maxwell models under study here should not be confused
with the
so-called Maxwell model (spring-damper) used as a minimal framework for the visco-elastic
properties in rheology. See e.g. DROZDOV A.D.,
\emph{Finite Elasticity and Viscoelasticity} (World Scientific) 1996.

\bibitem{BP}
BRILLIANTOV N. and P\"OSCHEL T.,
\emph{Kinetic Theory of Granular Gases} (Oxford University Press, Oxford) 2004.

\bibitem{S03}
SANTOS A., \emph{Physica A}, {\bf 321} (2003) 442; GARZ\'O V. and ASTILLERO A., \emph{J. Stat. Phys.}, {\bf 118} (2005) 935.

\bibitem{G03}
GARZ\'O V., \emph{J. Stat. Phys.}, {\bf 112} (2003) 657.

\bibitem{K05} KOHLSTEDT K., SNEZHKO A., SAKOZHNIKOV M.V., ARANSON I.S.,
OLAFSON J.S. and BEN-NAIM E., \emph{Phys. Rev. Lett.}, {\bf 95} (2005) 068001.

\bibitem{rque16}
As routinely done in the field, we define the
granular temperature kinetically, from the variance of the
velocity distribution \cite{IMM}. Such a quantity does not have any
thermodynamic basis. For equilibrium systems only
(no dissipation, no forcing) is it endowed with a thermodynamic
significance.

\bibitem{wp02}
WILDMAN R.D. and PARKER D.J., \emph{Phys. Rev. Lett.}, {\bf 88} (2002) 064301; FEITOSA K. and MENON N.,
\emph{Phys. Rev. Lett.}, {\bf 88} (2002) 198301.

\bibitem{GS03}GARZ\'O V. and SANTOS A., {\em Kinetic Theory of Gases in Shear Flows. Nonlinear Transport} (Kluwer Academic, Dordrecht) 2003.


\bibitem{NK02}
BEN-NAIM E. and KRAPIVSKY P. L., \emph{Eur. Phys. J. E}, {\bf 8} (2002) 507.

\bibitem{vanilla}MARINI BETTOLO MARCONI U. and PUGLISI A., \emph{Phys. Rev. E}, {\bf 65} (2002) 051305; {\bf 66} (2002) 011301; CONSTANTINI G., MARINI BETTOLO MARCONI U. and PUGLISI A., \emph{J. Stat. Mech.}, (2007) P08031.



\bibitem{GT10}
GARZ\'O V. and TRIZAC E.,
\emph{J. Non-Newtonian Fluid Mech.}, {\bf 165} (2010) 932.

\bibitem{GT11}
GARZ\'O V. and TRIZAC E., in preparation.

\bibitem{MSG96}
MAR\'IN C., SANTOS A. and GARZ\'O V., \emph{Europhys. Lett.}, {\bf 33} (1996) 599.

\bibitem{SD01}
SANTOS A. and DUFTY J. W., \emph{Phys. Rev. Lett.}, {\bf 86} (2001) 4823.

\bibitem{MG02}
MONTANERO J. M. and GARZ\'O V., \emph{Physica A}, {\bf 310} (2002) 17.



\bibitem{rque20}
Such a point of view leads to the Boltzmann equation
 for the excess component and the Lorentz-Boltzmann equation for the impurity and so, the impurity cannot retroact on the host medium.


\bibitem{Y71}YAMAKAWA H., \emph{Modern Theory of Polymer Solutions} (Harper and Row, New York, N.Y.) 1971.

\bibitem{RL77}R\'ESIBOIS P. and DE LEENER M., \emph{Classical Kinetic Theory} (Wiley, New York, N.Y.) 1977.


\end{thebibliography}
\end{document}